\documentclass[a4paper, 12pt]{article}

\usepackage[a4paper]{geometry}

\usepackage{url}
\usepackage{amsmath}

\usepackage{amssymb}
\usepackage{graphicx}
\usepackage{ctable}
\usepackage{threeparttable}
\usepackage{colortbl}
\usepackage{bigstrut}
\usepackage{here}
\usepackage{prettyref}
\usepackage{ntheorem}
\usepackage{arydshln}
\usepackage{subfloat}
\usepackage{subcaption}
\usepackage{dcolumn}

\usepackage{booktabs,mathptmx,siunitx}
\usepackage{hyperref}
\hypersetup{
    colorlinks = true,
    linkbordercolor = {white},
    citecolor =.,
    linkcolor =.,
   urlcolor = blue
}
\usepackage{setspace}
\usepackage{geometry}
\usepackage{bm}
\usepackage{natbib}
\usepackage{pxfonts}
\usepackage[multiple]{footmisc}

{\theorembodyfont{\normalfont}

}
  \theorembodyfont{\upshape}
  \theoremstyle{nonumberplain}
  \theoremseparator{}
  
\allowdisplaybreaks[4]
\begin{document}

\title{Spatial Pattern and City Size Distribution}

\author{Tomoya Mori\thanks{Institute of Economic Research, Kyoto University,
 Yoshida-honmachi, Sakyo-ku, Kyoto 606-8501, Japan. E-mail: mori@kier.kyoto-u.ac.jp}$\:\:^,$\thanks{Research Institute of Economy, Trade and Industry, 11th floor, Annex, Ministry of Economy, Trade and Industry (METI) 1-3-1, Kasumigaseki Chiyoda-ku, Tokyo, 100-8901 Japan.}}
\date{\today}
\maketitle 

\begin{abstract}
\noindent 
Many large cities are found at locations with certain first nature advantages. Yet, those exogenous locational features may not be the most potent forces governing the spatial pattern of cities. 
In particular, population size, spacing and industrial composition of cities exhibit simple, persistent and monotonic relationships. Theories of economic agglomeration suggest that this regularity is a consequence of interactions between endogenous agglomeration and dispersion forces.
This paper reviews the extant formal models that explain the spatial pattern together with the size distribution of cities, and discusses the remaining research questions to be answered in this literature. To obtain results about explicit spatial patterns of cities, a model needs to depart from the most popular two-region and systems-of-cities frameworks  in urban and regional economics in which there is no variation in interregional distance. This is one of the major reasons that only few formal models have been proposed in this literature. To draw implications as much as possible from the extant theories, this review involves extensive discussions on the behavior of the \emph{many-region extension} of these models. The mechanisms that link the spatial pattern of cities and the diversity in city sizes are also discussed in detail.
\end{abstract}
\noindent\textbf{Keywords:} {City size distribution, Spatial patterns, Agglomeration, Racetrack geography, Interregional distance, Power laws, Central place theory}
\bigskip

\noindent\textbf{JEL Classification:} R12, C33
\bigskip

\noindent \textbf{Acknowledgement: } The author thanks for the constructive and careful comments by two anonymous referees. This research was conducted as part of the project, “An empirical framework for studying spatial patterns and causal relationships of economic agglomeration,” undertaken at the Research Institute of Economy, Trade and Industry. This research has been partially supported by the Grant in Aid for Research (Nos. 17H00987, 16K13360, 16H03613,15H03344) of the MEXT, Japan. 
\thispagestyle{empty}
\setcounter{page}{0}

\newpage
\onehalfspacing
\section{Introduction}
In the past 50 years since the formal analyses of city formation started around the time of \cite{Alonso-1964},%
\footnote{There is a large literature on location theory that preceded urban economics and have important implications on city and agglomeration formation \citep[see, e.g.,][for a survey]{Thisse-et-al-Book1996}, although they were not designed to explain city formation per se.}\ %
the spatial pattern of cities has remained as a relatively minor subject in urban economics%
\footnote{A notable exceptions are \cite{Isard-QJE1949,Isard-Book1956}. While no formal models have been proposed by Isard, he foresaw the necessity of increasing returns and imperfect competition in order to explain the formation of cities and their spatial pattern. In particular, he envisaged the emergence of new economic geography which played a central role in this literature as will be discussed in Section \ref{sec:theory} \citep[see][for further discussions]{Fujita-JER2010}.}\ %
 -- despite that economic geographers in the past \citep[e.g.,][]{Thunen-1826, Christaller-1933,Losch-1940}, have commonly suggested the inseparable correspondence between (population) size and spatial distributions of cities \citep[see, e.g.,][]{Fujita-JER2010}.%
 \footnote{See an intriguing review by \cite{Fujita-RSUE2012} on the von Th\"{u}nen's work and ideas about spatial organization of economy.}\ %

The mainstream theories in urban economics have abstracted from the heterogeneity in inter-city/regional space by adopting \emph{the systems-of-cities model} since the pioneering work by J. Vernon Henderson \citep[][]{Henderson-1974},%
\footnote{See, e.g., \cite{Abdel-Rahman-Anas-HB2004} for a survey. See also  \cite{Behrens-Robert-Nicoud-HB2015} for more recent applications of this framework. 
The standard systems-of-cities models assume zero inter-city transport cost. There are a few variations assuming an equal distance between any pair of cities \citep[see, e.g.,][]{Anas-Xiong-JUE2003,Anas-JoEG2004}. In either case, there is only a single inter-city distance, as in the case of a two-region model.}\ %
or by simply assuming the presence of only two regions in an economy \citep*[see  a collection of two-region models presented in ][]{Baldwin-et-al-2003}.
The mechanism which determines the size of a city/region has always been a major subject in most of these theories, and for this purpose, the abstraction from interregional space in these approaches substantially simplified the analyses.

As a consequence of this particular evolution of the field, there exist rather limited theoretical as well as empirical literature which relate the spatial pattern and sizes of cities. To my knowledge, there are two major strands of formal models that explicitly deal with the spatial pattern of cities, \emph{new economic geography} (\emph{NEG}) and \emph{social-interactions models}.
The former explains city formation by the externalities that arise from monopolistically competitive markets \citep[see, e.g.,][\ for surveys]{Fujita-Krugman-Venables-1999,Baldwin-et-al-2003}, whereas the latter by the externalities that arise from direct interactions among agents outside the markets \citep[see, e.g.,][]{Beckmann-1976,Fujita-Ogawa-1982,Fujita-Smith-JRS1990,Mossay-Picard-JET2011}.
This paper focuses on the basic structure and implications of these theoretical models in connection to the observed size and spatial patterns of cities.

In Section \ref{sec:facts}, I start by making observations on the relation among sizes, spatial patterns and industrial structure of cities in reality by using data from Japan. 
In Section \ref{sec:theory}, generic properties of the canonical models (to be made precise below) of the extant theories are discussed. 
In particular, while most models were investigated in the context of the two-region setup in their original studies, in this paper, by extensively drawing from the work of \cite*{AMOT-DP2018}, I summarize their behavior in a many-region setup in which the spatial pattern of cities can be more properly studied. 
Finally, Section \ref{sec:conclusion} concludes the paper.

\section{Facts about size, location and industrial composition of cities\label{sec:facts}}

\noindent To guide summarizing and classifying the extant theoretical models for the size and spatial patterns of cities, it is useful to have a concrete idea about the basic relationships observed between them in reality. 
Given that the inter-city space has been largely abstracted in the literature, however, systematic researches on this subject are scarce, and the results published so far provide little decisive evidence \citep[e.g.,][]{Dobkins-Ioannides-2001, Overman-Ioannides-JUE2001, Ioannides-Overman-JoEG2004}. 
To demonstrate the strong correspondence between theories and facts, here, rather than trying to put together subtle pieces of evidences from the existing empirical literature, I attempt to develop a set of clear-cut facts using data from Japan.

There are two major reasons to focus on Japan as a real world example. One is the data availability of the micro data for industrial locations.
The other is the fact that both the highway and high-speed railway networks in Japan have been developed simultaneously almost from scratch to the full-fledged nation-wide networks between 1970 and 2015.
The changes in size and spatial patterns of Japanese cities in this period provides a useful test case to verify the implications from the theoretical models of endogenous city formation.
By utilizing these data, the key facts on all the aspects regarding size and spatial patterns as well as industrial structure of cities can be obtained for the same set of cities.

Throughout this section, a \emph{city} is defined to be a contiguous set of (approximately) 1km-by-1km cells with at least 1000 people per km$^2$ and total population of at least 10,000.%
\footnote{This definition of a city is a variation of that proposed by \cite{Rosenfeld-et-al-AER2011}.
The results to be presented below are not sensitive to the density and total population thresholds, unless they are set extremely high or low so that only a few high density cities or a few spatially gigantic cities are identified.}\ 
The advantage of this simple definition of a city is that the basic regional units (1km-by-1km cells) are consistent in the cross sections of a given country, and across different points in time, unlike more commonly used definitions of metropolitan areas based on administrative regions. Under this definition of a city, the set of all cities in a country account for the population (area) share in the country of 43.6\% (2.4\%), 44.6\% (1.6\%), 77.1\% (12.4\%), 48.7\% (2.9\%) and 47.0\% (3.8\%) for the US, Europe, Japan, China and India, respectively.%
\footnote{The estimated population count data at the 1km-by-1km cell level are obtained from \cite{PopCensus-2015} for Japan, and from the LandScan by \cite{LandScan-2015} for the rest of the countries.}\  

It is to be noted that the evidence on Japanese cities to be presented below is not specific to Japan.
For size and spatial patterns of cities discussed in Sections \ref{sec:spacing} and \ref{sec:size-dist}, \cite{Mori-et-al-DP2019} have shown that qualitatively the same (and more extensive) results hold for Japan and other five countries, China, France, Germany, India and the US under the same definition of a city. 
For the size and industrial structure of cities discussed in Section \ref{sec:hierarchy}, the qualitatively similar results have also been presented for the case of the US \citep[see][]{Hsu-EJ2012,Schiff-JoEG2014} under the standard metro areas and industrial classifications.

 \subsection{Size and spacing of cities\label{sec:spacing}}
 Many large cities are found at locations with certain first nature advantages.%
\footnote{For the role of the natural advantage in the city formation, see, e.g., \cite{Davis-Weinstein-AER2002} for the case of Japan, \cite{Bleakley-Lin-2012} and \cite{Cronon-1991} for the US, and \cite{Michaels-Rauch-EJ-2017} for France and the UK.}\ %
Yet, those exogenous locational features may not be the most potent forces governing the spatial pattern of cities. 
In particular, population size, spacing and industrial composition of cities exhibit a simple, persistent and monotonic relationship, which have long been recognized by economic geographers since \cite{Thunen-1826}, \cite{Christaller-1933} and \cite{Losch-1940}. 
They (especially, Christaller) suggested \emph{a central place pattern} in the relation between the size and location of cities such that \emph{larger cities tend to serve as centers around which smaller cities are grouped}. 
Moreover, this relation is recursive so that some of the small cities serve as centers around which even smaller cities are grouped. 
This central place pattern of cities naturally implies that \emph{a larger cities are more spaced apart}. 

\begin{figure}[H]
\centering{}\includegraphics[scale=0.375]{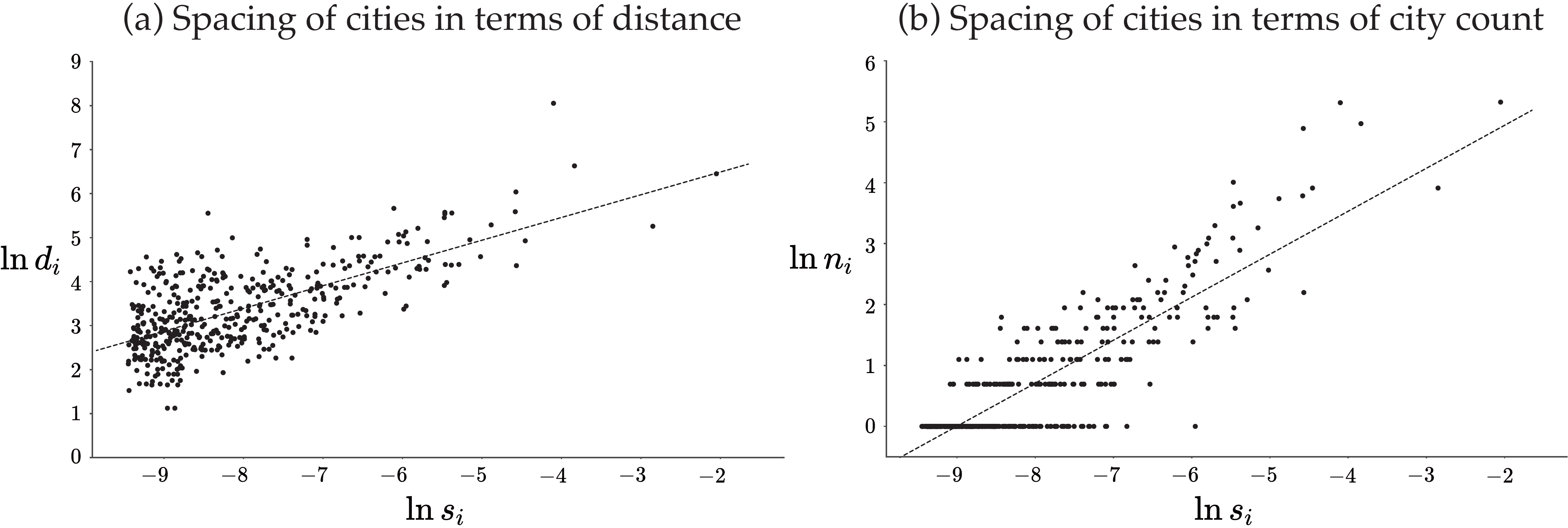}
\vspace{5pt}
 \caption{Spacing of Japanese cities in 2015\label{fig:spacing}}
\end{figure}

To see this in the actual data, let $\mathcal{U}$ be the set of all 450 cities identified in Japan in 2015, $s_i$ be the share of city $i\in \mathcal{U}$ in the national population, and $\|i,j\|$ for $i,j\in \mathcal{U}$ be the road distance between cities $i$ and $j$.%
\footnote{The road distance is based on the \href{https://www.openstreetmap.org/}{OpenStreetMap} data as of July, 2017. The distance between cities is computed as the distance between the centroids of the most densely populated 1km-by-1km cells in these cities. The computation was done using the Stata interface, osrmtime, of \href{http://project-osrm.org}{Open Source Routing Machine} by \cite{Huber-Rust-2016}.}\ 
Define the \emph{spacing} of city $i$ by the distance to the closest city of the same or a larger size class:%
\footnote{The lower threshold share, 0.75, defining the ``same size class'' in \eqref{eq:spacing} is arbitrary. But, the choice of the threshold value does not alter the qualitative result as long as it is not too far from 1.0.}\ 
\begin{equation}
	d_i = \min_{j\in \{k\in \mathcal{U}\;: \; s_k > 0.75s_i \}} \| i, j\|	\;. \label{eq:spacing}
\end{equation}
 Figure \ref{fig:spacing}(a) shows the relationship between $d_i$ and $s_i$ in log scale for each city $i\in \mathcal{U}$. The correlation between them is as high as 0.67.%
\footnote{The dashed line in the figure is the fitted line by Ordinary Least Squares (OLS) regression.}\ %
This confirms the spacing-out property of cities mentioned above.

If the number, $n_i$, of cities within the distance $d_i$ from city $i\in \mathcal{U}$ is counted by 
\begin{equation}
n_i \equiv \#\{j\in \mathcal{U}\backslash\{i\}\;:\; \|i,j\| < d_i\}\;,\label{eq:surrounding}
\end{equation}
as shown in Figure \ref{fig:spacing}(b), it also has strong correlation, 0.86, with the city size, $s_i$, in log scale. Thus, indeed it is clear that larger cities are surrounded by smaller cities.

\cite{Mori-et-al-DP2019} conduct a formal test for this central place pattern of cities, i.e., if  the largest cities are spaced out relative to the whole set of cities in a country.
Specifically, they first fix the number $L$ of the largest cities in a given country, and form a Voronoi partition with respect to a set of a given number $K\,(\geq 2)$ of randomly selected cities.
The test statistic is the count of partition cells containing at least one of these $L$ largest cities.
If there is substantial spacing between the largest cities in reality, then this count is expected to be larger for Voronoi partitions than for fully random partitions (i.e., without any regard to spatial relations among cities) of the same cell sizes.
For a range of $(L,K)$ values, they found strong evidence for the spacing-out of large cities in all the six countries considered (China, France, Germany, India, Japan and the US).%
\footnote{\cite{Dobkins-Ioannides-2001} found a negative correlation between the size and spacing of cities in the US for the period 1900-1980. But, the specific feature of the US cities needs to be taken into account is their historical development. The formation of cities started in the northeastern region of the US in the 19th century, and then expanded gradually to west and then to south. But, the \emph{effective} distance kept changing in the meantime in response to the advancement in the transport technology. 
As a consequence, the spacing of the same size class of cities has increased over time.
Such underlying heterogeneity across regions is to some extent taken into account in the construction of counterfactuals in the test by \cite{Mori-et-al-DP2019}.
}\ 	%

\subsection{Size distribution of cities\label{sec:size-dist}}

It is well known that city size distribution in a well-urbanized country exhibits an approximate power law \citep[e.g.,][]{Gabaix-Ioannides-2004, Batty-2006, Bettencourt-2013}.%
\footnote{\cite{Dittmar-DP2011} shows evidence that power laws for city size distributions in Europe emerged after 1500, i.e., after the dependence of city production on land relaxed substantially.}\ 
Formally, if a given set of $n$ cities is postulated to satisfy a power law, and if these city sizes are ranked as $s_1\geq s_2 \geq \cdots \geq s_n$, so that the rank, $r_i$, of city $i$ is given by $r_i = i$, then for some positive constants $c$ and $\alpha$,
\begin{equation}
	r_i / n \approx P(S>s_i) \approx cs_i^{-\alpha} \Rightarrow \ln s_i \approx b - \frac{1}{\alpha}\ln r_i
\end{equation}
for $b=\ln(cn) / \alpha$. 

\begin{figure}[H]
\centering{}\includegraphics[scale=.75]{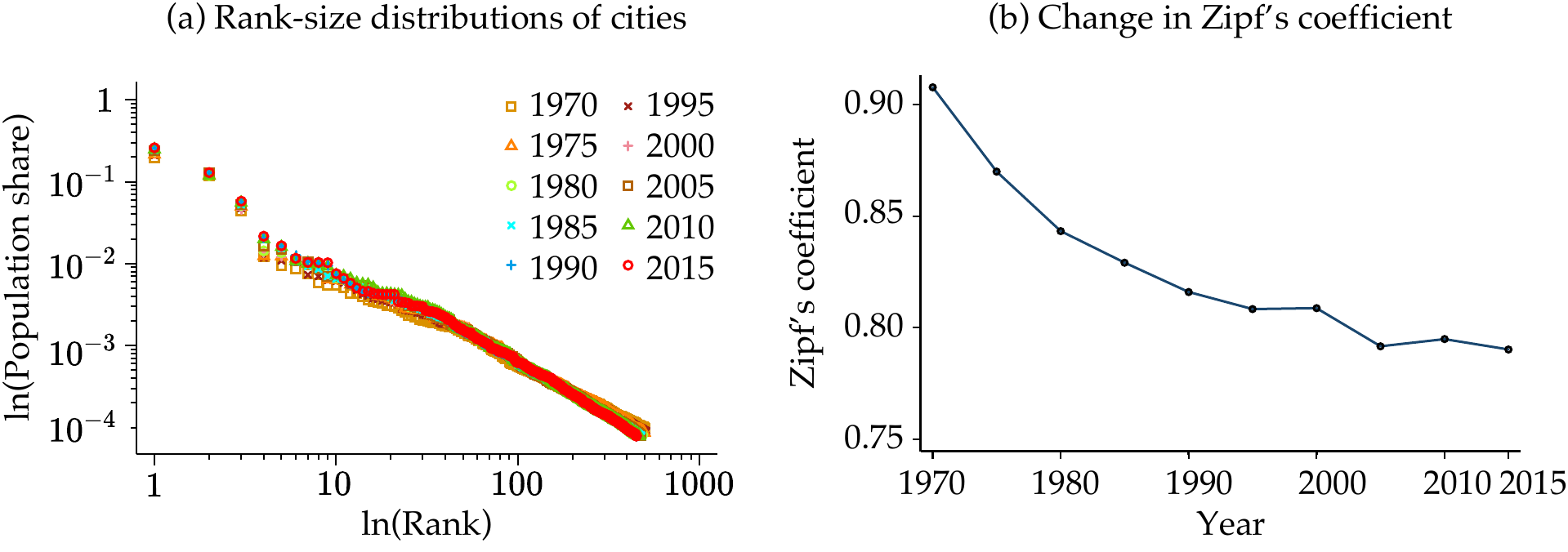}
\vspace{5pt}
 \caption{Rank-size distribution of cities in Japan\label{fig:rs}}
\end{figure}

In Figure \ref{fig:rs}, Panel (a) shows the rank-size distributions of cities in every five years from 1970 to 2015, where $s_i$ indicates the share of city $i$ in the national population; Panel (b) shows the change in $\alpha$, i.e., the Zipf's coefficient, over these 45 years. One can see that the city size distribution exhibits an approximate power law in each year, although agglomeration towards larger cities has been accelerated. The variation in city size is remarkably large, as exhibited by the largest three cities, Tokyo, Osaka and Nagoya, accounting for 45\% of the total city population, where Tokyo alone accounts for 26\% in 2015.

There is a strand of literature which informally argue that \emph{Zipf's law} \citep[after][]{Zipf-1949} holds, i.e., the power law with $\alpha=1$ holds for city size distribution in a country (see, e.g., \citeauthor{Gabaix-Ioannides-2004}, \citeyear{Gabaix-Ioannides-2004}; \citeauthor{Ioannides-Book2012}, \citeyear{Ioannides-Book2012}, \S8.2). But, there is abundant evidence against it \citep[e.g.,][]{Black-Henderson-JoEG2003,Soo-RSUE2005, Nitsch-JUE2005, Mori-et-al-DP2019}, as is also clear in the Japanese case shown in Figure \ref{fig:rs}.

 While the power laws for city size distributions are best known at the country level \citep[see, e.g.,][]{Gabaix-Ioannides-2004},  \cite{Mori-et-al-DP2019} have shown that the similar power laws appear recursively in spatial hierarchies of regions within a country that reflect the central place patterns discussed in Section \ref{sec:spacing}.
Specifically, they construct a spatial hierarchy in a country by first constructing a Voronoi partition of the set of all cities in the country using a given number of their largest cities as cell centers, and then continuing this partitioning procedure within each cell recursively. 
They found that city size distributions in different parts of these spatial hierarchies exhibit power laws that are significantly similar than would be expected by chance alone.
Thus, their result suggests that city systems have a spatial fractal structure within countries.

\subsection{Size and industrial structure of cities\label{sec:hierarchy}}
Many evidences \citep*[e.g.][]{Glaeser-Mare-2001, Bettencourt-et-al-2007, Combes-et-al-JUE2008, Glaeser-Resseger-2010, Combes-et-al-ECTA2012, Baum-Snow-Pavan-REStat2013, Davis-Dingel-DP2017} have indicated strong correlations between socio-economic quantities and sizes of cities (e.g., wages, education level, gross domestic product, industrial diversity, number of patents applications, amount of crime, level of traffic congestion). This section presents one of the clearest representations of such correlations by focusing on industrial location.

Let $\mathcal{I}$ be the set of all industries that operate in at least one of the cities, and for a given industry $i\in \mathcal{I}$, call a city \emph{a choice city} of this industry if industry $i$ is in operation in the city.
 These choice cities exhibit a systematic variation in their average population size across industries. 
  To see this, denote by $\mathcal{U}_{i}\: (\subseteq \mathcal{U})$ the set of all choice cities of industry $i\in \mathcal{I}$, then the average size of choice cities for industry $i$ is given by
\begin{equation}
	\overline{s}_i = \frac{1}{\#\mathcal{U}_i}\sum_{i\in \mathcal{U}_i} s_i\;,
\end{equation}
where $\#\mathcal{U}_i$ means the cardinality of set $\mathcal{U}_i$.

\begin{figure}[H]
\centering{}\includegraphics[scale=0.65]{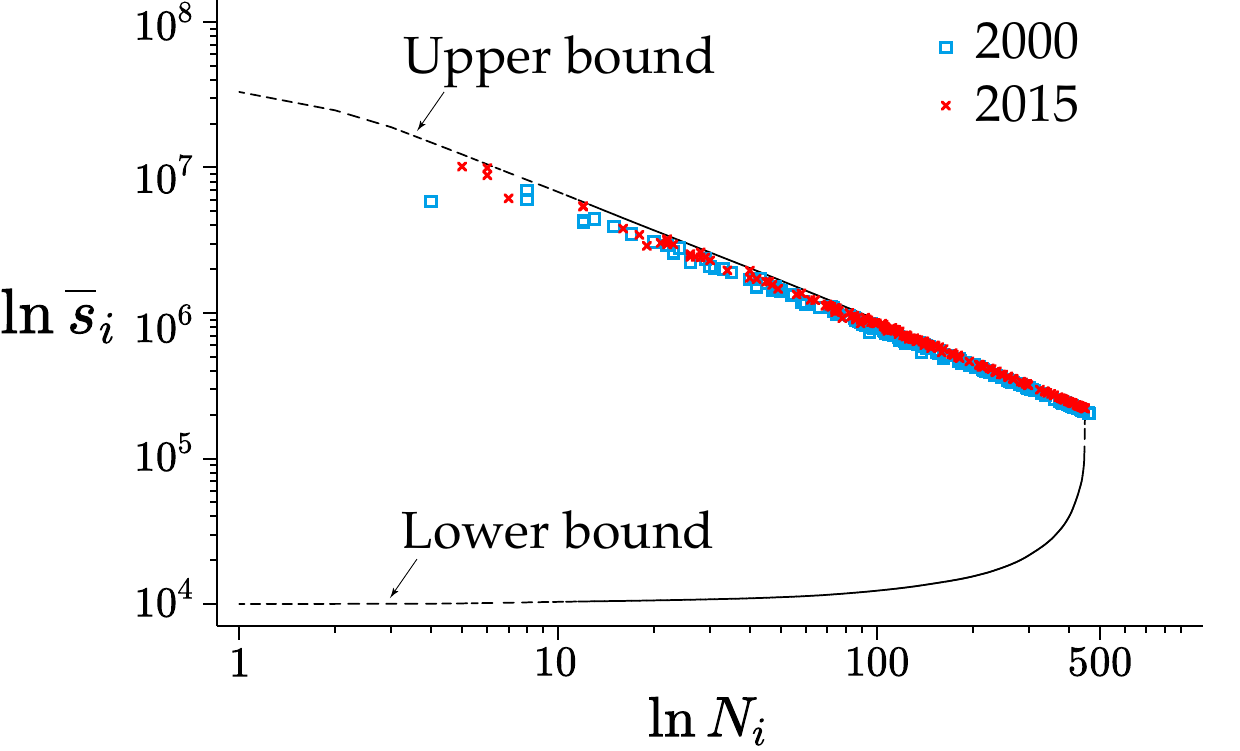}
\vspace{5pt}
 \caption{Varieties of economic activities and their choice of cities in Japan\label{fig:nas}}
\end{figure}

Now, consider three-digit secondary and tertiary industries of the Japanese Standard Industrial Classification (JSIC) that are present both in 2000 and 2015.  Of all the 237 such industries, there are 162 and 175 industries that have at least one establishment in cities in 2000 and 2015, respectively.%
\footnote{Data for the locations of establishments were obtained from \cite{EstabCensus-2000, EconCensus-2014}.}\ %
 Figure \ref{fig:nas} shows the relationship between $\overline{s}_i$ and $N_i$ for $i\in \mathcal{I}$  in log scale, where $N_i \equiv \#\mathcal{U}_i$. The dashed curves indicate the upper and lower bound for the average size of choice cities in 2015, where for each $i\in \mathcal{I}$, the former (latter) is the average size of the largest (smallest) $N_i$ cities.

There are two key features in these plots. First, the number $N_i$ and average size $\overline{s}_i$ of choice cities exhibit a strong \emph{power law}, which is persistent between 2000 and 2015.  Second, the average sizes of choice cities are almost hitting their upper bound, meaning that the choice cities of an industry $i\in \mathcal{I}$ is roughly the largest $N_i$ cities, which in turn implies that there is roughly a \emph{hierarchical relationship} in the industrial composition between larger and smaller cities.%
\footnote{These features are first recognized by \cite{Mori-et-al-JRS2008, Mori-Smith-JRS2011} for the case of Japan, and \citet[][Appendix A1]{Hsu-EJ2012} and \cite{Schiff-JoEG2014} for the case of the US. See also \cite{Davis-Dingel-DP2017} for an evidence of the hierarchical industrial structure of the US cities based on an alternative approach.}\ 

To see this, let $\mathcal{I}_i$ represent the set of industries that are present in city $i\in \mathcal{U}$, and for cities $i$ and $j\in \mathcal{U}$ such that $s_i>s_j$, define the \emph{hierarchy share} for city $j$ with $i$ by
\begin{equation}
	H_{ij} = 	\frac{\#\left(\mathcal{I}_i \cap \mathcal{I}_j\right)}{\#\mathcal{I}_j}\in [0,1]\;,
\end{equation}
where a larger value of $H_{ij}$ indicates a higher consistency with the hierarchical relationship, and  $H_{ij}=1$ means the perfect hierarchical relationship, i.e., $\mathcal{U}_j \subseteq \mathcal{U}_i$. The average values of the hierarchy shares for all the relevant city pairs,
\begin{equation}
	H \equiv  \frac{1}{\overline{H}}\sum_{i,j\in \mathcal{U}\:\:\text{s.t.}\:\: s_i > s_j} H_{ij}\:\:\in [0,1]
\end{equation}
where $\overline{H} \equiv \#\{(i,j)\::\: i,j\in\mathcal{U},\; s_i > s_j\}$, can be taken as an aggregate measure of spatial coordination among industries. A larger value of $H$ indicates a higher degree of spatial coordination, and the coordination is perfect if $H=1$. The actual values of $H$ are 0.76 and 0.80 in 2000 and 2015, respectively, which are quite high.\footnote{These values are much higher than the values of $H$ that can be realized under random location of industries after controlling for the industrial diversity (i.e., $\#\mathcal{I}_i$ for $i\in \mathcal{U}$) of cities and locational diversity (i.e., $\#\mathcal{U}_i$ for $i\in \mathcal{I}$) of industries \citep[see, e.g.,][]{Mori-et-al-JRS2008, Mori-Smith-JRS2011, Mori-ADR2017}.}\ %

 Together with the central place pattern discussed above (see Figure \ref{fig:spacing}), the fact that the spatial coordination of diverse economic activities leads to the diversity in city size has already been suggested informally by \cite{Christaller-1933} and \cite{Losch-1940}.

A large value of $H$ as in the case of Japan above means that it is not only that industries have different number of agglomerations (i.e., choice cities), but also that their locations tend to coincide, i.e., a more localized industry choose to locate in cities in which a more ubiquitous industries are present. The case of perfect coordination (i.e., $H=1$) corresponds to the \emph{hierarchy principle} in \cite{Christaller-1933}.
 
\begin{figure}[H]
\centering{}\includegraphics[scale=0.65]{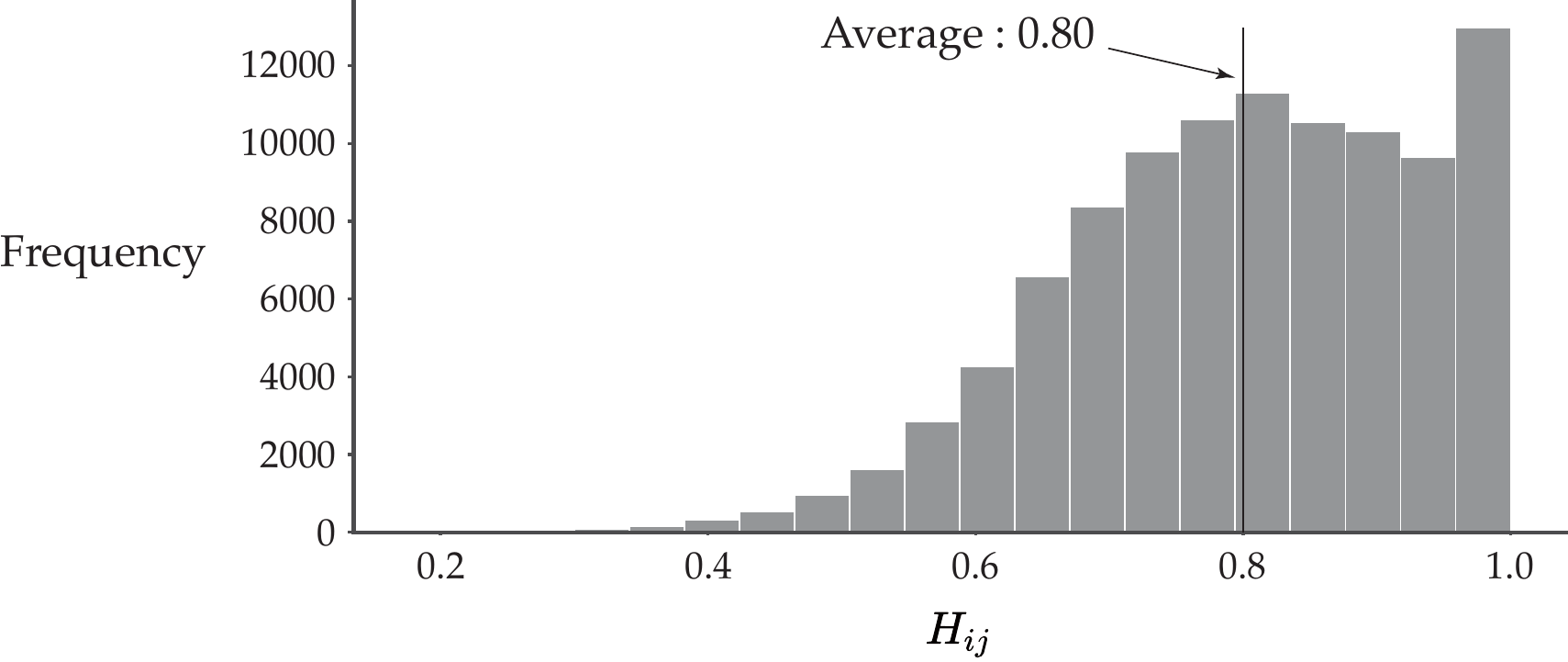}
\vspace{5pt}
 \caption{Distribution of hierarchy share between cities in Japan in 2015\label{fig:hshare}}
\end{figure}

To close this subsection, it is worth pointing out that while there is a strong tendency of hierarchical relation in the industrial composition between a larger and a smaller cities, it is by no means the rule. Figure \ref{fig:hshare} shows the distribution of $H_{ij}$ of all the relevant city pairs in 2015. While the mean value is $H=0.80$, the standard deviation is 0.13, and the range is from 0.18 to 1. Low hierarchy shares are realized for \emph{specialized cities} in which only a small specific set of industries are concentrated. 
As will be discussed in Section \ref{sec:sizes}, the standard systems-of-cities models \citep[e.g.,][]{Henderson-1974, Rossi-Hansberg-Wright-2007} associate the size of a city with that in scale economies specific to the industries in which the city is specialized, thereby explain the diversity in city size in terms of the variation in industry-specific scale economies.

\subsection{Growth of city sizes\label{sec:growth}}

Finally, we look at the characteristics of the growth of individual city sizes in Japan between 1970 and 2015. It is of particular interest to quantify the evolution of city sizes in this period, since it coincides with the period in which the highway and high-speed railway networks were developed almost from scratch to the extent that covers almost the entire nation, where the total highway (high-speed railway) length increased from 879 km (515 km) by more than 16 (10) times to 14,146 km (5,350 km).  

The level of interregional transport access has been one of the key parameters to determine the size and spatial patterns of cities in the literature. 
The evolution of the sizes of individual cities is expected to reflect the response to the improved interregional transport access, although the benefit for each city may vary depending on their relative location. Thus, the changes in size and spatial patterns experienced by Japanese cities in this period provides an ideal test case for the theoretical models of endogenous agglomeration.

There was substantial movement of population among cities in these 45 years.
In particular, there is a clear tendency of \emph{global agglomeration} toward a smaller number of cities, as the number of cities has decreased from 503 to 450.%
\footnote{Cities may emerge, disappear, split and merge over time. Cities identified in the consecutive two years are considered to represent the same city if they mutually account for the largest population among all the overlapping cities.}\ %

\begin{figure}[H]
\centering{}\includegraphics[scale=0.6]{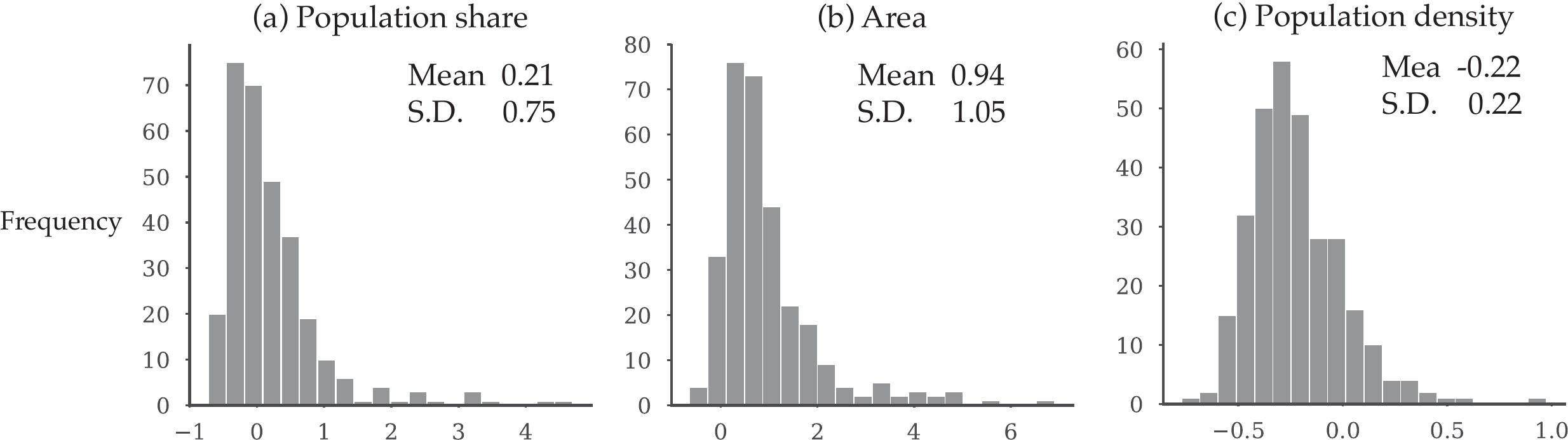}
\vspace{5pt}
 \caption{Changes in the sizes of individual cities in Japan between 1970 and 2015 (Growth rates in horizontarl axis)\label{fig:growth}}
\end{figure}

 Figure \ref{fig:growth} reveals key facts about the change in individual city sizes for the 302 cities that have remained throughout the entire period. Panel (a) adds another evidence for global agglomeration: the size of the remained cities in terms of population share (in the country) has grown by 21\% on average.%
\footnote{``S.D.'' in the panels means the standard deviation.}\ %
Note that it is more meaningful to look at the population share of a city rather than the population size itself to understand the tendency of global agglomeration, because the population shares remove the effects of general population growth and/or urbanization from the population sizes.%
\footnote{\cite{Overman-Ioannides-JUE2001} have shown evidence that there is mild tendency of the decrease in population size of relatively large cities (i.e., metropolitan areas with urban core of at least 50,000 population) of the US for the period 1920-1980. Their result is not directly comparable to the case of Japan here, since their results may be biased for relatively large cities, and the factors driving city sizes during the studied period were not made clear.}

Despite the tendency of global agglomeration, there is also a clear tendency of \emph{local dispersion} as the areal size of an individual city has almost doubled (Panel, b), while the population density has decreased by 22\% on average (Panel, c).%
\footnote{The suburbanization in response to the decrease in interregional transport access
 is one realization of local dispersion, and its evidence for the case of the US metro areas has been reported by \cite{Baum-Snow-QJE2007, Baum-Snow-DP2017}. For the global agglomeration and dispersion, no clear consensus has been attained at this point in the extant literature \citep[e.g.,][]{Duranton-Turner-REStud2012, Faber-REStud2014, Baum-Snow-DP2017}. This is rather evident from the discussion in Section \ref{sec:theory} below that the effect of interregional transport access on each individual city size is not monotonic. See \citet[][\S6]{AMOT-DP2018} for an extensive discussion on this respect. 
\cite{Ioannides-Overman-JoEG2004} examined the change in the distance from each city to its nearest neighbor, and found it was decreasing in the period of 1900 to 1990, which should essentially imply global dispersion. But, there is no discussion on the potential causes of this change in their paper.}\

This simultaneous occurrence of global agglomeration and local dispersion given an improvement in interregional access may seem paradoxical. But, it can be explained by integrating the extant theories of endogenous agglomeration to be discussed in the next section.%

\section{Theories\label{sec:theory}}
A model capable of explaining the spatial patterns of cities necessarily involves many regions with large variations in interregional distance, such that some cities are close to while others are far from one another. But, the majority of the extant models adopt either two-region%
\footnote{See, for example, \cite{Baldwin-et-al-2003} for a survey of NEG models}\ 
or systems-of-cities setups%
\footnote{See, for example, \cite{Anas-Xiong-JUE2003,Anas-JoEG2004,Tabuchi-Thisse-Zeng-JoEG2005}.}\ 
in which there is no variation in interregional distance.
Thus, no explicit spatial patterns reflecting the relation among the number, size and spacing of cities can be expressed by these models. 

A recent work by \cite{AMOT-DP2018} brought a breakthrough by showing that a wide variety of the extant models of endogenous agglomeration can be reformulated in a many-region setup, and formally analyzed in a unified framework. 
Specifically, they focus on a canonical model, i.e., a static model with (\emph{i}) a continuum of homogeneous mobile agents, each of whom chooses a single location; (\emph{ii}) there is a single type of interregional transport cost; (\emph{iii}) transport costs are subject to the iceberg technology.
The reformulated models are shown to boil down to one of the three reduced forms in terms of the spatial pattern of agglomeration and dispersion.

The canonical model covers a wide range of standard models in urban and regional economics.
It includes the class of NEG models based on the Dixit-Stiglitz type CES subutility function for love of variety \citep[e.g.,][]{Krugman-1991,Helpman-1998,Tabuchi-1998,Puga-EER1999,Forslid-Ottaviano-2003,Pfluger-2004,Murata-Thisse-2005,Redding-Sturm-AER2008, Pfluger-Suedekum-JUE2008}; social-interactions model of city-center formation based on technological externalities \citep[e.g.,][]{Beckmann-1976,Mossay-Picard-JET2011,Blanchet-et-al-IER2016}; and the economic geography models in ``universal gravity'' framework by  \cite{Allen-et-al-JPE2019}, including the \cite{Armington-IMF1969} model with labor mobility by \cite{Allen-Arkolakis-QJE2014}, a standard formulation in the recent quantitave spatial economics \citep[see, e.g.,][for a survey]{Redding-Rossi-Hansberg-ARE2017}.

Important classes of models that are out of their scope include city-center formation models in which firms and households have different location incentives (violation of (\emph{i})) \citep[e.g.,][]{Fujita-Ogawa-1982,Lucas-Rossi-Hansberg-ECTA2002,Ahlfeldt-et-al-ECTA2015,Monte-et-al-AER2018}; NEG models with additive transport costs (violation of (\emph{iii})) by \cite{Ottaviano-Tabuchi-Thisse-2002}, and those with multiple industries with industry-specific transport costs (violation of (\emph{ii})) \citep[e.g.,][]{Fujita-Krugman-1995, Fujita-Krugman-Mori-1999, Tabuchi-Thisse-JUE2011}. 


Drawing largely from \cite{AMOT-DP2018}, Section \ref{sec:canonical_model} reviews the mechanism underlying the relation between population/areal size and spacing of cities in reality discussed in Sections \ref{sec:spacing} and \ref{sec:growth}.
To explain the observed diversity in the size and industrial structure of cities discussed in Sections \ref{sec:size-dist} and \ref{sec:hierarchy}, respectively, and their relation to the spatial pattern of cities, a model needs to go beyond the canonical model considered by \cite{AMOT-DP2018}, and incorporate variations in the degree of increasing returns (and/or those in transport costs). At present, there are only a handful of models that have succeeded in such extensions. 
Section \ref{sec:sizes} reviews the theoretical development in this direction.

\subsection{Spatial pattern of cities\label{sec:canonical_model}}
By formalizing and generalizing the idea proposed by \citet[][Ch.8]{Krugman-Book1996} based on \cite{Turing-1952},  \cite*{Akamatsu-Takayama-Ikeda-2012} proposed an  analytical framework for many-region models of endogenous agglomeration under the symmetric racetrack geography with the help of discrete Fourier transformation.
While \cite{Akamatsu-Takayama-Ikeda-2012} has focused on a many-region extension of the model by \cite{Pfluger-2004}, \cite{AMOT-DP2018} have generalized their framework, and have shown that a wide variety of the extant models can be classified by the three distinct reduced forms, despite the difference in their specific mechanisms underlying agglomeration and dispersion. Below, I start by describing the basic setup of this approach.

\subsubsection*{Basic setup}
Consider the location space consisting of a set of $K$ discrete regions, $\mathcal{K}\equiv \{0,1,\ldots,K-1\}$. There is a continuum of homogeneous mobile agents whose regional distribution is denoted by $\bm{h}\equiv (h_i)_{i \in \mathcal{K}}$, where $h_i$ is the mass of mobile agents located in region $i$. 
Their total mass is a given constant, $H\equiv \sum_{i\in\mathcal{K}} h_i$. 
All regions in $\mathcal{K}$ are featureless and are placed at an equal interval on a circle.  
In this racetrack economy, transportation is possible only along the circumference.%
\footnote{The racetrack location space is obviously counterfactual, as it is edge less. Although the presence of the edge tends to make the agglomeration on the edge larger, since there is no competing agglomeration beyond the edge \citep[see, e.g.,][]{Fujita-Mori-RSUE1997, Ikeda-et-al-IJET2017}, this effect becomes negligible for a large economy, and the agglomeration patterns can be approximated by that in the edge-less economy.}\ 

Let region index $0,1,\ldots,K-1$ represent the location on the racetrack in clockwise direction.
Transport costs take iceberg form, i.e., if a unit of product  is shipped from region $i$ to $j$, then only the fraction $d_{ij} = d_{ji} \in [0,1)$ reaches $j$.
The \emph{spatial discounting matrix},  $\bm{D} = [d_{ij}]$, expresses the underlying distance structure of the economy. 
Typically, iceberg costs are expressed as $d_{ij} =\exp[-\tau\ell_{ij}]$, where $\ell_{ij}$ is the distance between regions $i$ and $j$ and $\tau\in (0,\infty)$ is the transport technology parameter.

The relocation of agents is assumed to be much slower than market reactions, so that the short-run equilibrium conditions
(such as market clearing and trade balance)
determine the payoff (utility or profit) in each region 
as a function of a given regional distribution of agents, $\bm{h}$.
Specifically, given $\bm{h}$, their \emph{short-run} payoff of choosing each region is determined, where
the short-run payoff function is denoted by $\bm{v}(\bm{h}) \equiv (v_i(\bm{h}))_{i\in \mathcal{K}}$, with $v_i(\bm{h})$ representing the payoff for an agent located in region $i\in \mathcal{K}$.

In the \emph{long-run}, agents are mobile 
and are free to choose their locations
to improve their own payoffs. 
In (\emph{long-run}) \emph{equilibrium}, it must hold that $v^* = v_i(\bm{h})$
for all regions $i$ with $h_i > 0$, and
$v^* \geq v_i(\bm{h})$
for any region $i$ with $h_i = 0$, where $v^*$ is the equilibrium payoff level.

A change in endogenous agglomeration pattern is treated as an instance of bifurcation of an equilibrium. 
To address the stability of equilibria, a standard approach in the literature
is to introduce equilibrium refinement based on \emph{local stability} 
under myopic evolutionary dynamics, where 
the rate of change in the number of residents $h_i$ 
in region $i$ is modeled 
on the basis of the regional distribution of agents, $\bm{h}$, 
and that of payoff, $\bm{v}(\bm{h})$.
Let a deterministic dynamic be denoted by 
$\dot{\bm{h}} = \bm{F}(\bm{h},\bm{v}(\bm{h}))$,
where $\dot{\bm{h}}$ represents the time derivative of $\bm{h}$, and assume that 
(i) 
$\bm{F}$ satisfies
differentiability with respect to both arguments, 
(ii) agents relocate in the direction 
of an increased aggregate payoff under $\bm{F}$, 
(iii) the total mass of agents is preserved under $\bm{F}$, and 
(iv) any spatial equilibrium is a rest point of the dynamic, i.e., if $\bm{h}^*$ is an equilibrium,
	it must hold that $\dot{\bm{h}} = \bm{F}(\bm{h}^*,\bm{v}(\bm{h}^*)) = \bm{0}$.
 The stability of an equilibrium then is defined in terms of asymptotic stability under $\bm{F}$.
\subsubsection*{Formation of a city}
With a racetrack geography, the uniform distribution of mobile agents is always an equilibrium when the payoff function is symmetric across regions. Call an equilibrium with uniform distribution a \emph{flat-earth equilibrium}, and denote it by
$\bar{\bm{h}} \equiv (h, h, \hdots, h)$ 
with $h \equiv H/K$.

If the adjustment dynamic is formulated so that the agents migrate in order to maximize their payoff, it follows \citep[Appendix B]{AMOT-DP2018} that each eigenvalue of Jacobian matrix $\bm{J}$ of $\bm{F}$ and that of the Jacobian matrix $\nabla \bm{v}$ of $\bm{v}$ are real, and have a perfect positive correlation at the flat-earth equilibrium.
Thus, one can focus on $\nabla \bm{v}$ instead of $\bm{J}$ to investigate the stability of the flat-earth equiliburum.
What remains is to identify the direction of the bifurcation at the flat-earth equilibrium, which is equivalent to find the eigenvector of $\nabla \bm{v}(\bar{\bm{h}})$ whose eigenvalue changes its sign from negative to positive first among all the eigenvectors of $\nabla \bm{v}(\bar{\bm{h}})$.

The sign of the $k$-th eigenvalue of $\nabla \bm{v}(\bar{\bm{h}})$ has been shown to coincide with the sign of the model-specific function of the form:
\begin{equation}
		G\left(f_k\right)=c_0+c_1 f_k+	c_2 f_k^2, 	\label{eq:G-f}
\end{equation}
where $c_0, c_1$ and $c_2$ are the constants specific to a given model, and $f_k$ is the $k$-th eigenvalue of the spatial discounting matrix $\bm{D}$ which is known to be real, and common to all models. The eigenvector associated with $f_k$ is given by $\bm{\eta}_k = (\eta_{k,i}) = (\cos[\theta k i])$ for $i\in \mathcal{K}$ with $\theta \equiv 2\pi/K$, and the bifurcation from the flat-earth equilibrium takes place in the direction given by $\bm{h} = \bar{\bm{h}} + \epsilon \bm{\eta}_k$ with $\epsilon > 0$.

\begin{figure}[h!]
\centering{}\includegraphics[scale=0.65]{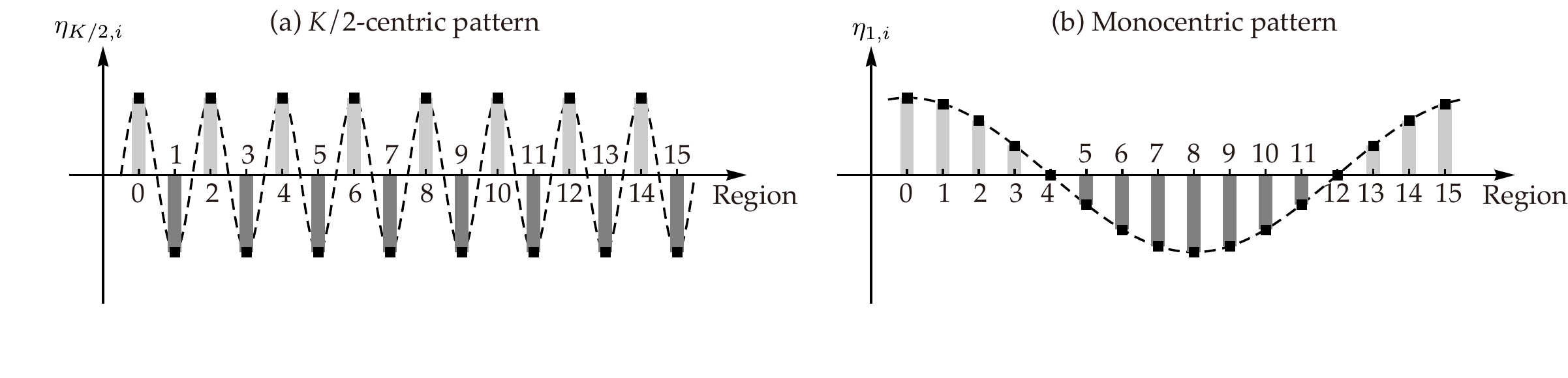}
\vspace{5pt}
\caption{Agglomeration formation at high and low transport costs\label{fig:eigenvectors}\protect\footnotemark}
\end{figure}

The value $k$ coincides with the number of equidistant regions toward which mobile agents migrate the most.
For example, at $k=K/2$, the value $\eta_{K/2,i}$ of each element $i\in\mathcal{K}$ in eigenvetor,  $\bm{\eta}_{K/2}$, is given as depicted for the case of $K=16$ in Figure \ref{fig:eigenvectors}(a), so that agglomerations start to form at alternate regions, $0,2,4,\ldots,K-2\,(=14)$.%
\footnote{It is equally likely that agglomerations take place at regions, $1,3,\ldots,K-1$.}\ 
At $k=1$, as depicted in Figure \ref{fig:eigenvectors}(b), an unimodal agglomeration will form around region 0.%
\footnote{It is equally likely that the agglomeration takes place around any region in $\mathcal{K}$.}\ 
\footnotetext{This figure is the replication of \citet[][Figure 3]{AMOT-DP2018}.}

There are two key properties of $f_k's$ that are useful to investigate the stability of flat-earth equilibrium:
\begin{enumerate}
	\item $f_k$ is monotonically increasing in transport cost, $\tau$.
	\item $f_1 =\max_{k= 1,2,\ldots,K} f_k$ and $f_{K/2} = \min_{k=1,2,\ldots, K}  f_k > 0$.\footnote{$f_0$ whose corresponding eigenvector is $\eta_0 = (1,1,\ldots,1)$ is irrelevant for the stability of equilibria as the total mobile population is preserved.}\footnote{For simplicity, it is assumed that $K$ is an even integer, although it is not essential.}
\end{enumerate}

Canonical models typically have a positive value of $c_1$. Since $f_1 > 0$, it means that the second term on the right hand side (RHS) in \eqref{eq:G-f} represents the agglomeration force, as it works to destabilize the flat-earth equilibrium.
In these models, if a stable flat-earth equilibrium exists, then one must have either $c_0 <0$ or $c_2 < 0$, or both, so that all the eigenvalues of $\nabla \bm{v} (\bar{\bm{h}})$ can be negative at the flat-earth equilibrium. 
In particular, since $f_k$ is positive and increasing in $\tau$ for each $k=1,2,\ldots,K-1$, the flat-earth equilibrium is stable for sufficiently small transport costs if $c_0 < 0$,  and for sufficiently large transport costs if $c_2 < 0$.

The bifurcation from the flat-earth equilibrium leading to the city formation under $c_0 <0$ and that under $c_2 < 0$ are, however,  qualitatively different in two aspects. 
The first aspect is the timing at which the bifurcation takes place. 
The bifurcation under $c_0<0$ takes place in the increasing phase of transport costs, whereas that under $c_2<0$ in their decreasing phase.

\begin{figure}[h!]
\centering{}\includegraphics[scale=0.8]{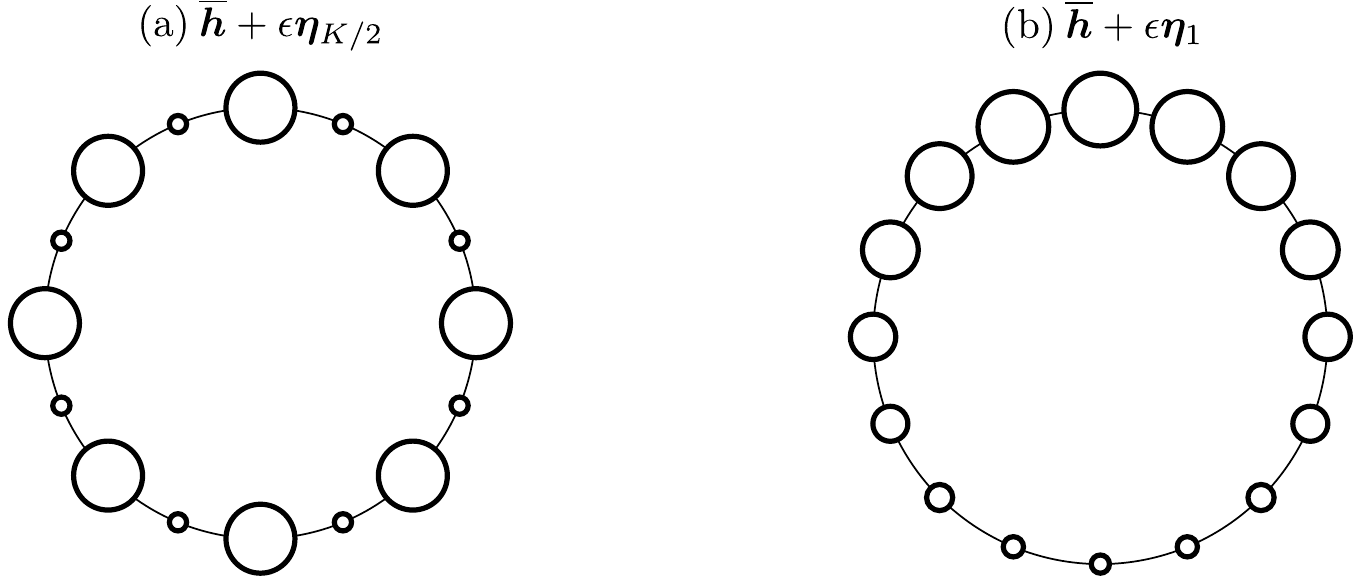}
\vspace{5pt}
\caption{City formation at high and low transport costs\label{fig:first_bifurcation}\protect\footnotemark}
\end{figure}

The second aspect is the spatial scale of agglomeration and dispersion. 
Provided that $c_2 <0$, the bifurcation takes place in the direction of $\eta_{K/2}$, i.e., every other region along the racetrack attracts in-migration of mobile agents, when $G(f_{K/2})$ becomes positive (refer to Figure \ref{fig:eigenvectors}(a)). 
The regional distribution of mobile agents that arises in this bifurcation is $\bar{\bm{h}}+\epsilon \bm{\eta}_{K/2}$ (for $\epsilon > 0$) as illustrated in Figure \ref{fig:first_bifurcation}(a).
In other words, small cities (i.e., agglomerations) form locally, while they are dispersed globally all over the location space.
 
Provided that $c_0<0$, the bifurcation takes place in the direction of $\bm{\eta}_1$ when $G(f_1)$ turns to positive (refer to Figure \ref{fig:eigenvectors}(b)).
The regional distribution of mobile agents that arises in this case is given by $\bar{\bm{h}}+\epsilon \bm{\eta}_1$ as illustrated in Figure \ref{fig:first_bifurcation}(b).
In other words, the agglomeration takes place globally, and form a single gigantic city, while the dispersion takes place locally around the city center (region 0) so that the city stretches over the entire location space.
\footnotetext{This figure is the replication of \citet[][Figure 5]{AMOT-DP2018}.}

A crucial difference between the two cases is the dependence of dispersion force on the distance structure of the model.
The third term $c_2f_k^2$ on the RHS of \eqref{eq:G-f} \emph{depends} on the distance structure of the economy (through $f_k$). 
As discussed above, this force leads to \emph{global dispersion} (with local agglomeration) as in Figure \ref{fig:first_bifurcation}(a).
On the contrary, the first term on the RHS of \eqref{eq:G-f} is the dispersion force when $c_0 < 0$ which is \emph{independent} of the distance structure of the economy.
As discussed above, this force leads to \emph{local dispersion} (with global agglomeration) as in Figure \ref{fig:first_bifurcation}(b).

\begin{figure}[H]
\centering{}\includegraphics[scale=0.8]{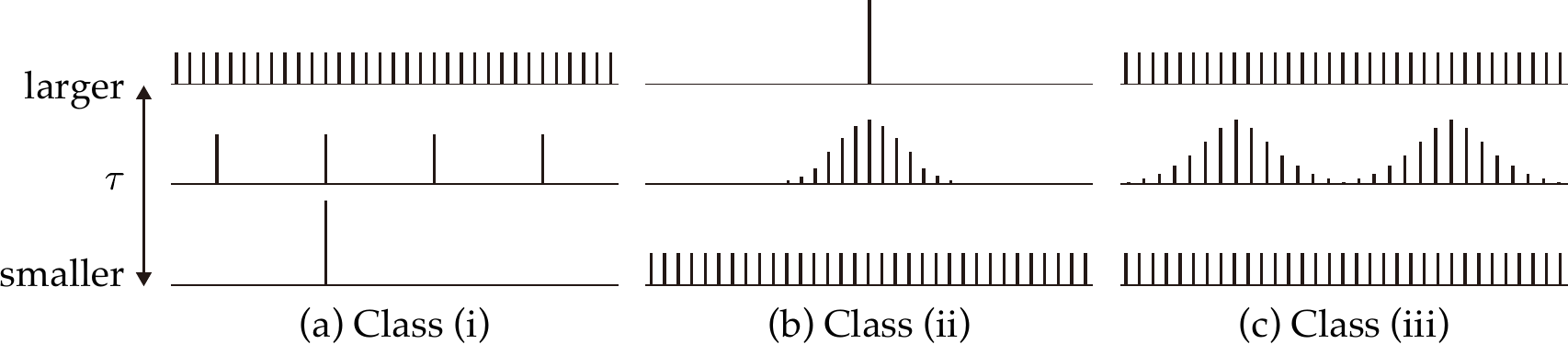}
\vspace{5pt}
 \caption{Spatial patterns of cities\label{fig:patterns}}
\end{figure}

In \cite{AMOT-DP2018}, the models with only global dispersion force, i.e., $c_0\geq 0$ and $c_2 <0$, are called Class (i). 
The models of this class are shown to exhibit \emph{period doubling bifurcations} as transport costs decrease, leading to \emph{a smaller number of larger cities with a larger spacing between neighboring cities, until all mobile agents concentrate in one region} (Figure \ref{fig:patterns}a).%
\footnote{See \cite{Akamatsu-Takayama-Ikeda-2012} for the formal analyses on the period doubling bifurcations of class (i) models.}\ 
 The models with only local dispersion force, i.e., $c_0<0$ and $c_2\geq 0$ are called Class (ii). 
 The Class (ii) models involve at most one bifurcation when the flat-earth equilibrium looses stability. 
 In the models of this class, \emph{keeping unimodal regional distribution, the concentration of mobile agents proceeds as transport costs increases, until all mobile agents concentrate in one region}  (Figure \ref{fig:patterns}b).
The models that incorporate both types of dispersion force, i.e., $c_0<0$ and $c_2<0$, may be the most realistic, and account for the formation of multiple cities with a positive internal space (Figure \ref{fig:patterns}c). These are called Class (iii). 

Two implications are worth mentioning. First, the heterogeneity among interregional distances is an essential feature of a model to investigate the spatial pattern of cities. In the context of a two region model or a systems-of-cities model in which there is no variation in interregional distance, the dispersion of mobile agents in Class (i) and Class (ii) models look exactly the same. But, as indicated by the middle panels of Figure \ref{fig:patterns}(a)(b), these are qualitatively different in spatial scale. The dispersion takes place at the global scale in Class (i) models -- in the form of an increase in the number of cities, and at the local scale in Class (ii) models -- in the form of a larger spatial extent of a city.

Second,  the responses of agglomeration/dispersion to the level of transport costs are \emph{opposite} between global and local spatial scales. 
More specifically, given the lower interregional transport costs, the agglomeration proceeds at global scale, i.e., the number of cities decreases, the sizes and the spacing of the remaining cities increase, while the dispersion proceeds at local scale, i.e., the average population density within a city decreases and the spatial extent of a city increases.%
\footnote{Of course, the actual evolution of the spatial patterns under the changing level of transport costs is more complicated, as neighboring cities may eventually merge in the case of Class (iii) models. See, \citet[][\S 5.3]{AMOT-DP2018}.}\ %

Notice that the behavior of Class (i) models essentially account for the larger cities being spaced more apart as discussed in Section \ref{sec:spacing}, and the behavior of Class (iii) models, i.e., the combination of Classes (i) and (ii), can account for the  evolution of city growth of Japan discussed in Section \ref{sec:growth}.

\medskip

Below, I overview a variety of extant models that fall into one of these three classes, as well as those do not.

\subsubsection*{New economic geography}
NEG \citep*[e.g.,][]{Fujita-Krugman-Venables-1999} commonly utilizes the monopolistic competition together with scale economies in production to explain the endogenous agglomeration. On the one hand, the love for product variety by consumers and the presence of transport costs give an incentive for consumers to locate closer to firms. On the other hand, each indivisible firm subject to scale economies at the plant level has an incentive to locate and supply near the concentration of consumers.%
\footnote{An alternative formulation assumes the product variety in intermediate goods. See, e.g., \citet[][Ch.14]{Fujita-Krugman-Venables-1999}.}

In this context, the global dispersion force associated with $c_2 < 0$ in \eqref{eq:G-f} is introduced typically by assuming immobile consumers in each region who generate dispersed demand for the differentiated products \citep[e.g.,][]{Krugman-1991, Krugman-1993, Forslid-Ottaviano-2003, Pfluger-2004}. The assumption of immobility of consumers is nothing but simplification to assure the dispersed demand. It can be obtained endogenously, for example, by introducing land-intensive sectors that also require labor inputs \citep[e.g.,][]{Fujita-Krugman-1995, Puga-EER1999}, which in turn generates dispersed demand from workers. With transport costs, the proximity to demand matters, and hence, the spatial dispersion of consumers results in the formation of multiple cities, where the firms in each city mainly serves their nearby local market.

The local dispersion force associated with $c_0<0$ in \eqref{eq:G-f} is introduced by assuming consumption of locally scarce land \citep[e.g.,][]{Helpman-1998, Redding-Sturm-AER2008, Redding-Rossi-Hansberg-ARE2017}, sometimes together with commuting costs \citep[e.g.,][]{Murata-Thisse-2005}.%
\footnote{A similar effect can be obtained by assuming local congestion externality that is effective within a given region.}\ %
All these costs of concentration are confined within a given region, and thus are independent of interregional distance. 
The dispersion in this case takes the form of overflow of mobile agents from a given city to the nearby regions, rather than the formation of new distinct cities at distant regions.

There are models that incorporate both global and local dispersion forces above \citep[][]{Tabuchi-1998, Pfluger-Suedekum-JUE2008}, i.e., of Class (iii) with $c_0<0$ and $c_2<0$ in \eqref{eq:G-f}. While these themselves treat only the two-region case, their many-region extensions can generate a more realistic spatial pattern of cities that involve both global and local dispersion as shown in Figure \ref{fig:patterns}(c) \citep[see][\S5.3]{AMOT-DP2018}.%
\footnote{NEG models adopting transport costs that are not iceberg form are not studied in \cite{AMOT-DP2018}. But, it is known that they can also be classified according to the spatial scale of dispersion. For example, \cite{Ottaviano-Tabuchi-Thisse-2002} and \cite{Tabuchi-Thisse-Zeng-2005}, both of which adopt additive transport costs, belong essentially to Class (i) and Class (ii), respectively \citep[see][\S3.1]{AMOT-DP2018}.}\ %

\subsubsection*{Social interactions model}

In the 1970s and 1980s, there were a series of attempts to explain endogenous formation of the central business districts (CBD) \emph{within a city}. The development of the models of this type was initiated by \cite{Solow-Vickrey-JET1971} and \cite{Beckmann-1976}, then followed by several others \citep[e.g.,][]{Borukhov-Hochman-EPA1977,OHara-JPE1977, Ogawa-Fujita-1980, Fujita-Ogawa-1982, Imai-1982, Tauchen-Witte-1983, Tabuchi-RSUE1986, Fujita-1988,Kanemoto-1990,  Fujita-Book1990}. 

In these models, the formation of CBD is explained by introducing positive technological externalities generated from the interaction between each pair of individual agents. 
While the above mentioned models vary in the specification of positive externalities, \cite{Fujita-Smith-JRS1990} have shown that their formulations are essentially equivalent, and reformulated commonly by the so-called \emph{additive interaction function}, $S_i (\bm{h}) \equiv  \sum_{j\in \mathcal{K}} d_{ij} h_j$.

In the simplest specifications \citep[as in, e.g.,][]{Beckmann-1976}, this additive interaction function enters the utility function of consumers directly. Most models assume land consumption by mobile agents, while the production sector is abstracted, i.e., they incorporate only local dispersion force, and hence belong to Class (ii). One exception is \cite{Takayama-Akamatsu-JSCE2011} who also included global dispersion force by introducing mobile firms and immobile consumers in each region. This model thus contains both local and global dispersion force, i.e., of Class (iii).%
\footnote{The social interactions model by \cite{Picard-Tabuchi-JET2013} with non-iceberg transport costs can be shown to belong to Class (iii) \citep[see][\S3.1]{AMOT-DP2018}.}\

\subsubsection*{Other relevant models}
In the NEG literature, a particularly important deviation from the canonical models is to consider different transport cost structures by industry. For example, \cite{Fujita-Krugman-1995} included transport costs for (urban) differentiated products as well as land-intensive (rural) homogenous products. 
In the presence of rural goods that are costly to transport, the delivered price for such goods is lower in regions farther away from cities, which generates a dispersion force. This is similar to the local dispersion force in that even a small deviation from an urban agglomeration will decrease the price of rural goods and increase the payoff of the deviant. However, the advantage of dispersion persists outside the agglomeration, i.e., it depends on the distance structure of the model. This type of dispersion force has been shown to result in the formation of an \emph{industrial belt}, a continuum of agglomeration associated with multiple atoms of agglomeration as demonstrated by the simulations in \cite{Mori-JUE1997} and \cite*{Ikeda-et-al-IJET2017}. The formal characterization of industrial belts, however, remains to be carried out. 


Among the extant social-interactions models, some distinguish location incentives between firms and consumers/workers unlike the canonical models discussed above \citep[e.g.,][]{Ogawa-Fujita-1980, Fujita-Ogawa-1982, Ota-Fujita-RSUE1993, Lucas-Rossi-Hansberg-ECTA2002}. 
This distinction is especially crucial for explaining the location patterns within a city, while it may be less relevant for the purpose of explaining the spatial pattern of cities.
At present, little formal results have been obtained regarding the spatial pattern of cities that arise in these models (see \citeauthor{Osawa-2016}, \citeyear{Osawa-2016}, for the recent theoretical development in this direction.)

Other relevant models that were not covered so far include the \emph{spatial oligopoly} models designed to explain the agglomeration of retail stores \citep[e.g.,][]{Wolinsky-Bell1983, Dudey-AER1990, Konishi-JUE2005}.
 In these models, consumers have imperfect information on the types and prices of goods sold by stores before they visit them. The greater the agglomeration of stores, the more likely it is that consumers will find their favorite commodities. The concentration of stores is explained by the market-size effect due to taste uncertainty and/or lower price expectation.
 The dispersion force is global one given by the exogenous and spatially dispersed demand. Thus, these models are expected to behave similarly to Class (i) models above, although no extensive analyses have been conducted in this direction \citep[see][\S5, for the discussion on the spacing of retail clusters]{Konishi-JUE2005}.%
 \footnote{See \cite{Economides-Siow-AER1988} for a related model that explains the spacing of market areas in which markets are formed due to matching externalities that arise in the exchange of consumption goods.}

\subsection{Diversity in city size\label{sec:sizes}}

The most popular theoretical explanation of power law for city-size distribution at this point may be the \emph{random growth theory} (e.g., \citeauthor{Gabaix-1999}, \citeyear{Gabaix-1999}; \citeauthor{Duranton-2006}, \citeyear{Duranton-2006}, \citeyear{Duranton-AER2007}; \citeauthor{Rossi-Hansberg-Wright-2007}, \citeyear{Rossi-Hansberg-Wright-2007}; \citeauthor{Ioannides-Book2012}, \S8.2, \citeyear{Ioannides-Book2012}) which postulates that the growth rates of individual cities follows Gibrat's law \citep[][]{Gibrat-Book1931}, i.e., they are independently and identically distributed random variables.

This theory is highly compatible with systems-of-cities models. 
For example, in the model by \cite{Rossi-Hansberg-Wright-2007}, individual industries are subject to city-level positive externality from agglomeration, but do not benefit from colocation with other industries, so that the externality is industry-specific. Then, each city would specialize in a single industry in the presence of urban costs due to scarcity of land and the need for commuting in a city.
If the industry- (or city-) specific productivity growth rates satisfy the basic assumptions of random growth theory (including Gibrat's law), the model generates the power law for city size distributions.
It is a plausible explanation, as we have seen in Section \ref{sec:hierarchy} that the specialized cities are rather ubiquitous (refer to Figure \ref{fig:hshare} and the corresponding discussion) despite the strong evidence for the hierarchy principle \emph{\`{a} la} \cite{Christaller-1933}.

A key implication of the random growth theory is that similar power laws hold for all (sufficiently large) random subsets of cities in a country,  i.e., \emph{without any regard to spatial relation among cities.}
Thus, this theory essentially denies the mutual dependence of size and spatial patterns of cities.
But, as discussed in Section \ref{sec:size-dist}, \cite{Mori-et-al-DP2019} have shown that the similarity in power laws for city size distributions is much stronger among the cells in the spatial hierarchical partitions of cities that are consistent with the central place patterns than among random subsets of cities, i.e., if city sizes were generated by a random growth process.%
\footnote{See \cite{Rosenfeld-et-al-2008,Rybski-Ros-PHYR-E2013} for other evidence against Gibrat's law for city sizes.}\footnote{There still are possibilities to extend random growth models by adding spatial relations among cities, thereby account for the spatial fractal structure of city systems in terms of power laws of city size distributions. See, for example, \citet[][\S8.2.5]{Ioannides-Book2012} for a review of related attempts.}

To account for the large diversity in city size observed in reality by the many-region models described in Section \ref{sec:canonical_model}, one needs to incorporate diversity in increasing returns (and/or that in transport costs).
While Class (i) models with a global dispersion force discussed above can account for the formation of multiple cities, there is little variation in the sizes of cities to be realized in equilibrium, since each model has only one type of increasing returns. 

There has been attempts of formalizing the central place thoery of \cite{Christaller-1933} by introducing multiple industries subject to different degrees of increasing returns.
The initial formal attempt was made by \cite{Beckmann-1958}. But, his model lacked microeconomic foundation. Later the models with more explicit mechanisms were developed by \cite*{Fujita-Krugman-Mori-1999,Tabuchi-Thisse-JUE2011} in the context of the NEG, and by \cite{Hsu-EJ2012} in the context of spatial competition model. 
In these models, the different degrees of increasing returns among industries result in the different spatial frequencies of agglomeration among industries.

The key to account for the diversity in city size in these models is the \emph{spatial coordination} of agglomerations among industries through inter-industry demand externalities that arise from common consumers among industries. An industry subject to a larger increasing returns agglomerate in a smaller number of cities that are farther apart. What is crucial is that these cities are chosen from the ones in which more ubiquitous industries subject to smaller increasing returns are located.
Consequently, larger cities are formed at the location in which the coordination of a larger number of industries takes place.
This spatial coordination of industries accounts for the positive correlation between the size, spacing and industrial diversity of a city as observed in reality (Sections \ref{sec:spacing} and \ref{sec:hierarchy}).

In particular, \cite{Hsu-EJ2012} proposed a unique spatial competition model with product differentiation and scale economies in production, and provided at this point the most far reaching formal explanation for the mutual dependence between spatial pattern and size diversity of cities. 
When the distribution of scale economies in production of each firm (which is expressed in terms of the industry-specific fixed cost for production in his model) is \emph{regularly varying}, then his model replicates the power law for city size distribution (Section \ref{sec:size-dist}) together with the positive correlation between size and spacing of cities (Section \ref{sec:spacing}), the power law for the number and the average size of choice cities of industries (Section \ref{sec:hierarchy}), as well as the hierarchy principle observed in Japan (Section \ref{sec:hierarchy}).

\cite{Davis-Dingel-DP2017} offer an alternative mechanism of spatial coordination among industries which in turn results in hierarchy principle and the diversity in city sizes in the context of a systems-of-cities model.%
\footnote{\cite{Rossi-Hansberg-Wright-2007} formulate a random-growth model by using the systems-of-cities model in which cities are specialized in a single different industries, and power laws emerge for sizes of these cities.}\ 
Specifically, the hierarchy principle in this model arises from vertical heterogeneity in skill level among workers and skill requirement by industries together with inter-industry positive externality that is confined within the same city.
The mechanism underling the spatial coordination among industries in this model is different from the central place models above.
On the one hand, a small city attracts only low skill industries and  workers as it offers only small city-level agglomeration externality.
On the other hand, a large city attracts both high and low skill industries and workers.
High skilled have an incentive to live there, since the city offers a large city-level agglomeration externality and they can afford to live there.
Although residential locations near the city center are occupied by high skilled, low skilled still can afford to live in locations with low land rent (due to longer commuting) near the city fringe, while enjoying the large city-level externality.

Alternatively, \cite{Desmet-Rossi-Hansberg-JET2009}, \cite{Desmet-Rossi-Hansberg-AER2014}, \cite{Desmet-Rossi-Hansberg-JRS2015}, \cite{Desmet-et-al-JPE2017} 
incorporated dynamic externalities through endogenous innovation and spillover effects. 
These models are fundamentally different from all the models discussed so far in that the exogenous heterogeneity among regions are essential for city formation, i.e., agglomerations do not form spontaneously. 
The uneven distribution of mobile agents resulting from the exogenous regional heterogeneity is magnified by the spillover effects over time.
One exception in this strand of literature is \cite{Nagy-DP2016} who incorporated the same dynamic externalities into the NEG framework, so that his model is capable of explaining the spontaneous formation of multiple cities together with the diversity in city sizes. While this model has been applied to replicate the evolution of the US cities in 19th century, the properties of agglomeration and dispersion in this model have not been formally analyzed.

\section{Concluding remarks\label{sec:conclusion}}
This paper reviewed the models which explain the mutual dependence of spatial pattern and sizes of cities. A many-region geography with variations in interregional distance is an essential feature of a model, if the spatial pattern of cities were the subject of the study. 
Naturally, there have been very few formal attempts that explicitly dealt with this high-dimensional problem until recently with notable exceptions by \cite{Hsu-EJ2012}.

A breakthrough has been brought about by \cite{Akamatsu-Takayama-Ikeda-2012} who proposed to focus on the racetrack economy which involves many regions with heterogeneous interregional distances, while preserving symmetry among the regions. By utilizing the discrete Fourier transformation, they have demonstrated that the spatial patterns of agglomeration that aries in the NEG models in a many region setup can be formally analyzed to a large extent. The same group of researchers have also developed the framework for systematic numerical analysis on a many-region geography based on the \emph{numerical bifurcation theory} and \emph{group-theoretic bifucation theory}  (e.g., \citeauthor*{Ikeda-Akamatsu-Kono-2012}, \citeyear{Ikeda-Akamatsu-Kono-2012}; \citeauthor{Ikeda-et-al-IJET2017}, \citeyear{Ikeda-et-al-IJET2017}). Their numerical approach makes it possible to explore asymmetric geography (e.g., the presence of edges and heterogeneity in regional advantages) as well as two-dimensional location space in a many-region setup.

In this paper, drawing largely from \cite{AMOT-DP2018} which applied the analytical tool developed by \cite{Akamatsu-Takayama-Ikeda-2012} to a wide variety of extant agglomeration models, I have reviewed the spatial pattern of cities and its relation to city sizes implied by these models. But, \cite{Hsu-EJ2012} continues to be the only tractable model that can account for the large diversity in city size in association with the observed spatial pattern of cities. Thus, much to be expected in the future development in this respect.

Finally, no models so far have been successful in integrating intra- and inter-city space. In the models aiming to explain intra-city spatial patterns, the location behavior of firms and that of workers are typically distinguished, and land consumption and/or land inputs by firms together with commuting are considered \citep[e.g.,][]{Fujita-Ogawa-1982, Ota-Fujita-RSUE1993, Lucas-Rossi-Hansberg-ECTA2002, Picard-Tabuchi-JET2013}. The models aiming to explain inter-city spatial patterns, on the contrary, typically ignore different location incentives between firms and workers (all models discussed in this paper belong to this group). 
But, it is not trivial to integrate these two spatial scales in one model.

Some extant NEG models consider commuting and land consumption \citep[e.g.,][]{Anas-JoEG2004, Murata-Thisse-2005}. But, such urban structure is by assumption confined within a given region, and does not extend beyond a single region.
As is discussed in Section \ref{sec:canonical_model}, in a many-region geography with variations in interregional distance, these models belong to Class (ii), which means that at most  unimodal agglomeration forms. 
Although each region in these models has monocentric urban structure \emph{by assumption}, and hence, it is tempted to be interpreted as a ``city'', they can generate essentially at most one ``true'' city.

To fully account for the spatial pattern of cities, the distinction between inside and outside each city should also be endogenized.

\newpage
\singlespacing
\bibliographystyle{aer} 
\bibliography{SpatialPattern.bib}

\end{document}